\documentclass[preprint,showpacs,preprintnumbers,amsmath,amssymb]{revtex4}

\usepackage{graphicx}
\usepackage{dcolumn}
\usepackage{bm}

\begin{document}

\preprint{APS/123-QED}

\title{Roles of non-equilibrium conduction electrons on magnetization
dynamics of ferromagnets} 

\author{S. Zhang and Z. Li}

\affiliation{Department of Physics and Astronomy, University of
Missouri-Columbia, Columbia, MO 65211} 

\date{\today}

\begin{abstract}
The mutual dependence of spin-dependent conduction and magnetization
dynamics of ferromagnets provides the key mechanisms in various spin-dependent
phenomena. We compute the response of the conduction electron spins to 
a spatial and time varying magnetization ${\bf M} ({\bf r},t)$ within the 
time-dependent semiclassical transport theory.
We show that the induced non-equilibrium conduction spin
density in turn generates 
four spin torques acting on the magnetization--with each torque playing
different roles in magnetization dynamics. By 
comparing with recent theoretical models, we find that one of these torques 
that has not been previously identified is crucial to
consistently interpret experimental data on domain wall motion.
\end{abstract}


\maketitle

Recently, there are emerging interests in the interplay between 
spin-dependent transport properties and magnetization dynamics of ferromagnets.
Giant magnetoresistive effect in magnetic multilayers \cite{Baibich}
is one of the examples that the spin-transport is controlled by magnetization 
dynamics (or configurations). Spin angular momentum transfer \cite{Slon}, 
or spin torque, manifests the magnetization dynamics controlled by 
spin-polarized conduction electrons. There are quite a few closely 
related phenomena reported recently, e.g., enhancement of damping parameters 
due to spin pumping \cite{Bauer,Heinrich} and reaction spin torques
\cite{Ho}, dynamic RKKY interaction
\cite{Heinrich2}, spin echo \cite{Bauer2} and adiabatic spin torques 
in a domain wall \cite{Tatara}. 
These proposed or observed phenomena motivated us to look for a theoretical
framework which is capable to address the above phenomena on an equal footing. 
The essence of the above phenomena is to recognize two types of electrons:
spin-dependent transport is provided by electrons at the Fermi level and
magnetization dynamics may involve electrons below 
the Fermi sea. While it is impossible to unambiguously 
separate electrons of transport from electrons of magnetization 
in a real ferromagnet, it has been conventionally modeled via a ``s-d'' 
Hamiltonian,
\begin{equation}
H_{sd} = - {J_{ex}}{\bf s}\cdot {\bf S}
\end{equation}
where ${\bf s}$ and ${\bf S}$ are the 
spins of itinerant and localized electrons,
and $J_{ex}$ is the exchange coupling strength.
In this letter, we show that the above simple s-d model in fact captures 
most of the physics on the interplay between spin-polarized 
transport of itinerant electrons and magnetization dynamics of local 
moments. We will first derive a linear response function
for the conduction electron spin in the presence of 
a time and spatially varying local moment, and then by using the same s-d 
model to calculate the spin torque on the magnetization dynamics as a result
of the induced non-equilibrium conduction electron spin. Among other things, 
we have found four distinct
spin torques on the magnetization. Three of them are closely related to
previously derived torques by using different methods. One of the 
derived torque is new; it describes the mis-tracking between the conduction 
electron spin and the spatially varying local moment. 
We further show that our formulation can be conveniently applied to study 
magnetization dynamics. An example of domain wall motion is illustrated in the
end of the paper.

The dynamics of the conduction electron will be considered separately from that
of local magnetization. We treat the itinerant spin ${\bf s}$ as a full 
quantum mechanical operator whose 
equation of motion is governed by a transport equation, but we approximate 
${\bf S}$ as a classical magnetization vector whose dynamics is much slower
than that of itinerant spins, i.e., we replace ${\bf S}$ by
a classical magnetization ${\bf M} ({\bf r}, t)$
\begin{equation}
H_{sd} = - \frac{J_{ex}}{\hbar M_s} {\bf s}\cdot {\bf M}({\bf r},t)
\end{equation}
where $|{\bf M}({\bf r},t)|=M_s$ is 
the saturation magnetization. We first determine
the induced spin density for a given ${\bf M} ({\bf r},t)$ and then derive
the reaction of the induced spin density to the magnetization.

In the present study, the non-equilibrium conduction electrons are
generated by applying either a DC electric field or a time-dependent 
magnetic field. While the electric field directly generates the charge 
and spin currents in conducting ferromagnets, the time-dependent
magnetic field is to drive the magnetization motion that induces a
non-equilibrium spin density via ``s-d'' interaction. 
The conduction electron spin operator satisfies the generalized
spin continuity equation, 
\begin{equation}
\frac{\partial {\bf s}}{\partial t} + \mbox{\boldmath $\nabla$} \cdot
\hat{J} = \frac{1}{i\hbar} [{\bf s}, H_{sd}] - \Gamma_{re}({\bf s})
\end{equation}
where $\hat{J}$ is the spin current operator, and $\Gamma_{re} ({\bf s})$ 
represents the spin relaxation due to scattering with impurities, electrons,
etc. By defining electron spin density ${\bf m} ({\bf r},t) = <{\bf s}> $ 
and spin current density ${\cal J} ({\bf r},t) = <\hat{J}> $ 
where $<>$ represents the average over all occupied electronic states,
e.g., $ <{\bf s}> = {\rm Tr} (\rho {\bf s})$ where the trace is over all 
electronic as well as spin states, and $\rho $ is the density operator, 
one obtains a semiclassical Bloch equation for the conduction 
electron spin density,
\begin{equation}
\frac{\partial {\bf m} }{\partial t} + 
\mbox{\boldmath $\nabla$} \cdot {\cal J} 
= - \frac{1}{\tau_{ex} M_s} {\bf m} \times {\bf M}({\bf r},t)
- <\Gamma({\bf s})> 
\end{equation} 
where the commutator in Eq.~(3) has been explicitly calculated by utilizing
Eq.~(2), and we have defined $\tau_{ex} = \hbar/J_{ex}$. 

Next, we separate the induced spin density ${\bf m}$ into two terms,
\begin{equation}
{\bf m}  ({\bf r},t) = {\bf m}_0 ({\bf r},t) + \delta {\bf m}  ({\bf r},t)
=n_0 \frac{{\bf M} ({\bf r},t)}{M_s} +  \delta {\bf m}  ({\bf r},t)
\end{equation}
where $n_0$ is the local equilibrium spin density whose direction
is parallel to the magnetization. The first term in Eq.~(5) represents 
the equilibrium spin density when the conduction
electron spin relaxes to its equilibrium value at an instantaneous time $t$.
Since the dynamics of the magnetization is slow compared to that of
conduction electrons, it is reasonable to assume 
the spin of the conduction electrons approximately follows the direction 
of the local moment, known as the adiabatic process. The second term
represents the deviation from this adiabatic process. Similarly, we
write the spin current density as  
\begin{equation}
{\cal J} ({\bf r},t) = {\cal J}_0 ({\bf r},t) + \delta {\cal J} ({\bf r},t)
=-(\mu_BP/e) {\bf j}_e \otimes
\frac{{\bf M} ({\bf r},t)}{M_s} + \delta {\cal J} ({\bf r},t)
\end{equation}
where $e$ is the electron charge, $j_e$ is the current density,
$\mu_B$ is the Bohr magneton, and
$P$ is the spin current polarization of the ferromagnet. 
Note that the spin current is a tensor that consists 
of two vectors: the charge current and the spin polarization of the current.
The first term in Eq.~(6) is
the spin current whose spin polarization  
is parallel to the local magnetization ${\bf M} ({\bf r},t)$. To solve for 
the non-equilibrium spin density in a closed form, we consider the following
simplifications. First, we use a simple relaxation time approximation to
model the relaxation term in Eq.~(4), i.e., 
we write $<\Gamma({\bf s})>= \delta {\bf m} ({\bf r},t) / \tau_{sf}$ where
$\tau_{sf}$ is the spin-flip relaxation time. The approximation is necessary
in order to obtain a simple analytic expression. 
Second, we only consider the linear response of $\delta {\bf m}$ to
the electric current $j_e$ and to the time derivative of magnetization 
$\partial {\bf M}/\partial t$. Since $\delta {\bf m}$ is already the first 
order, $\partial \delta {\bf m}/\partial t $ will be the order of
$j_e \cdot \partial {\bf M}/\partial t$ 
or $ \partial^2 {\bf M}/\partial t^2$ and thus it can be discarded. 
Within the semiclassical picture of the transport, the 
non-adiabatic current density $\delta {\cal J}$ is related to 
the non-equilibrium spin density $\delta {\bf m}$ via $\delta {\cal J}
= -D_0 \mbox{\boldmath $\nabla$} \delta {\bf m}$ where $D_0$ is 
the diffusion constant. 
By inserting Eqs.~(5) and (6) into (4) and utilizing the above simplification,
we obtain the closed form for the non-equilibrium spin density 
\begin{equation}
D_0 \nabla^2 \delta {\bf m} - \frac{1}{\tau_{ex} M_s} \delta {\bf m} 
\times {\bf M} - \frac{\delta {\bf m}}{\tau_{sf}} 
= \frac{n_0}{M_s} \cdot \frac{{\partial \bf M}}{\partial t}
- \frac{\mu_BP}{eM_s} ({\bf j}_e \cdot \mbox{\boldmath $\nabla$}) {\bf M}
\end{equation}
One immediately realizes that the non-equilibrium spin density is created
by two source terms on the right side of Eq.~(7): one is the time 
variation and the other is the spatial variation of the magnetization. 
The solution of the above differential equation depends on the detail
structure of the magnetization vector. Here we assume that the magnetization
varies slowly in space, i.e., the domain wall width $W$ of the magnetization
is much larger than the transport length scale defined in the  
footnote [9]. In this case, the spatial derivation, the first term 
in Eq.~(7), can
be discarded \cite{footnote2}. Then Eq.~(7) becomes a simple vector algebraic
equation and by using the elementary vector manipulation 
we readily obtain an explicit expression for 
the non-equilibrium spin density
\begin{equation}
\delta {\bf m} = 
\frac{\tau_{ex}}{(1+\xi^2)} \left(
- \frac{\xi n_0 }{M_s} \frac{\partial {\bf M}}{\partial t}
- \frac{n_0}{M_s^2} {\bf M} \times \frac{\partial {\bf M}}{\partial t}
+ \frac{\mu_BP\xi}{eM_s}( {\bf j}_e \cdot \mbox{\boldmath $\nabla$}) {\bf M} 
+ \frac{\mu_BP}{eM_s^2}
{\bf M} \times ({\bf j}_e \cdot \mbox{\boldmath $\nabla$}) {\bf M} \right)  
\end{equation}
where $\xi = \tau_{ex}/\tau_{sf}$.
The above induced spin density in turn exerts a spin torque on the
magnetization. From Eq.~(2), the torque is 
${\bf T}= -(J_{ex}/\hbar M_s) {\bf M} \times {\bf m} = 
-(J_{ex}/\hbar M_s) {\bf M} \times \delta {\bf m}$.
By using Eq.~(8), we have
\begin{equation}
{\bf T}= \frac{1}{1+\xi^2} \left(
- \frac{n_0}{M_s} \frac{\partial {\bf M}}{\partial t}
+ \frac{\xi n_0 }{M_s^2} {\bf M} \times \frac{\partial {\bf M}}{\partial t}
-  \frac{\mu_BP}{eM_s^3} {\bf M} \times [ {\bf M}\times 
({\bf j}_e \cdot \mbox{\boldmath $\nabla$})  {\bf M}]
- \frac{\mu_BP\xi }{eM_s^2} {\bf M} \times 
({\bf j}_e \cdot \mbox{\boldmath $\nabla$}) {\bf M} \right)
\end{equation}
There are four terms; the first two are from magnetization variation
in time and the last two in space. Interestingly, the first
two terms are independent of the current. 
The last two terms represent the 
current-driven effect since they are proportional
to the current. We now discuss the role of each spin torque below. 

The standard Landau-Lifshitz-Gilbert (LLG)
equation consists of a precessional term
due to an effective field and a phenomenological damping term. In addition
to these two torques, the above torque ${\bf T}$ is now added to 
the LLG equation,
\begin{equation}
\frac{\partial {\bf M}}{d t} = - \gamma {\bf M} \times
{\bf H}_{eff} + \frac{\alpha}{M_s} \left(
{\bf M} \times \frac{\partial \bf M}{d t} \right) + {\bf T}
\end{equation}
where $\gamma$ is the gyromagnetic ratio, ${\bf H}_{eff}$ is the effective 
magnetic field, $\alpha$ is the Gilbert damping parameter. 
We immediately realize that the first term in Eq.~(9) is simply to
renormalize the gyromagnetic ratio while the second term is to
renormalize the damping parameter. Thus if we introduce an 
effective gyromagnetic ratio $\gamma'$ and the damping parameter
$\alpha'$, 
\[
\gamma' = \gamma (1+\eta )^{-1}; \hspace{0.3in} \gamma ' \alpha' = 
\gamma (\alpha+\xi \eta)
\]
where we have defined $\eta = (n_0/M_s)/(1+\xi^2)$, LLG equation remains 
in the same form.
We point out that the modification of the gyromagnetic ratio and
the damping parameter through the present 
mechanism is rather small in transition metal ferromagnets. For a 
typical ferromagnet (Ni, Co, Fe and their alloys), 
$J_{ex} \approx 1$ eV, $\tau_{sf} \approx 10^{-12}$s, $n_0/M_s 
\approx 10^{-2}$, $\xi \approx 10^{-2}$ 
and thus $\eta$ is about $10^{-2}$ and $\xi \eta$ 
is of the order of $10^{-4}$--much smaller than the typical
damping parameter of the order of $10^{-2}$.  
Therefore, we conclude that the temporal spin torque driven
by the exchange interaction only slightly modifies the damping parameter
and can not be identified as a leading mechanism for magnetization damping. 

At this point, we should compare other theories on the spin torque.
Tserkovnyak {\em et al.} \cite{Bauer,Bauer2} proposed an adiabatic spin pumping 
mechanism to explain the enhancement of Gilbert damping parameters. 
Ho et al suggested a radiation field induced by magnetization 
precessional motion of magnets \cite{Ho}. Most recently, a similar s-d model 
in the presence of the time-dependent magnetization has been considered
\cite{Tser}. The present approach reduces to these theories in 
the simple limit considered for these two terms. 
In fact, the idea of this temporal spin torque had been suggested 
earlier: when the magnetization varies in time, the spin of the conduction 
electrons tends to follow the direction of the magnetization with a time delay
given by spin relaxation time; this phenomenon was named as 
``breathing Fermi surface'' \cite{Kambersky}. We are now able to
consider this physics of the enhanced damping on the equal footing as
the current induced spin torques. 

Our main focus here is the spin torque due to the spatially non-uniform 
magnetization vector, the last two terms in Eq.~(9). 
Since the temporal spin torques can be completely absorbed by the
re-definition of the gyromagnetic ratio and damping constant, we should
now just ignore them and concentrate on the role of spin torque generated
by the non-uniform magnetization.
We thus write the full equation for the magnetization dynamics below
\begin{equation}
\label{LLG} \frac{\partial {\bf M}}{\partial t}=-\gamma {\bf
M}\times {\bf H}_{eff}+\frac{\alpha}{M_s} {\bf M} \times \frac{\partial
{\bf M}}{\partial t}-\frac{b_{J}}{M_s^2} {\bf M}\times\left({\bf
M}\times \frac{\partial{\bf M}}{\partial x}
\right)-\frac{c_{J}}{M_s}{\bf M}\times \frac{\partial {\bf M}}{\partial
x}
\end{equation}
where we assume the direction of current $x$-direction 
(${\bf j}_e =j_e {\bf e}_x$), $b_J = Pj_e \mu_B /eM_s(1+\xi^2)$, 
and $c_J =Pj_e \mu_B \xi /eM_s(1+\xi^2)$. 
Note that $b_J$ and $c_J$ have the unit of velocity. 
The ``$b_J$'' term has been already proposed by Bazaliy et al. \cite{Bazaliy}
when they consider a ballistic motion of conduction electrons in the
half-metal materials. Recently Tatara and Kohno also derived similar
expression \cite{Tatara}. We have seen that this term describes 
the adiabatic process of the non-equilibrium conduction electrons.
The ``$c_J$'' term is completely new; it is
related to the spatial mis-tracking of spins between conduction
electrons and local magnetization. While this term is known in 
the physics of domain wall resistance \cite{Levy,Tatara2,Simanek}, 
it also gives rise a non-adiabatic spin torque, the last term in Eq.~(11). 
At first sight, one might think that this ``$c_J$'' term may be discarded
since it is much smaller than the ``$b_J$'' term ($c_J/b_J = 
\xi \approx  10^{-2}$). We will show below 
that the terminal velocity of a domain wall 
is {\em independent} of the strength of ``$b_J$'', rather it is 
controlled by this small ``$c_J$'' term. Thus, experimental analysis 
on the domain wall motion must include this new $``c_J''$ term.

To make a concrete prediction on the domain wall dynamics from Eq.~(11), we 
consider a N\'{e}el wall in a magnetic nanowire whose magnetization vector 
only depends on the position along the wire, i.e., ${\bf M}={\bf M}(x,t)$. 
The effective field entering
Eq.~(11) is modeled by 
\begin{equation}
{\bf H}_{eff} = \frac{H_K M_x}{M_s} {\bf e}_x + \frac{2A}{M_s^2}
\nabla^2 {\bf M} - 4 \pi M_z {\bf e}_z + H_{ext} {\bf e}_x
\end{equation}
where $H_K$ is the anisotropy field, $A$ is the exchange constant,
and $4 \pi M_z$ is the de-magnetization field. 
In the presence of the spin torque, we
follow the Walker's prescription of the domain wall motion
by introducing a trial function ${\bf M} (\theta, \varphi)$ 
where $(\theta, \phi)$ are polar angles in the following form \cite{Walker}, 
\begin{eqnarray}
\varphi=\varphi({\it t}); \hspace{0.1in}
\ln\tan\frac{\pi-\theta}{2}= \frac{1}{W(t)}\left(x-\int_{0}^{\it t}{\it
v}(\tau) d \tau\right) \,
\end{eqnarray}
The first equation assumes that the projection of the
magnetization vector in the domain wall on the $yz$ plane is
independent of the position. The second equation in Eq.~(13) 
postulates that the domain
wall shape remains a standard N\'{e}el-wall form except that the
wall width $W(t)$ varies with time
and the wall moves at velocity ${\it v}({\it
t})$. By placing Eqs.~(13) and (12) 
into Eq.~(11), and by assuming the domain wall width changes slowly as
in the Walker's theory, we can find 
two coupled differential equations for determining the domain wall distortion
parameters $\varphi (t)$ and $W(t)$. Interestingly, the expression for the 
velocity of the domain wall at the initial application of the current is 
\cite{footnote}
\begin{equation}
v(0) = - \frac{1}{1+\alpha^2} \left(
\frac{\alpha \gamma H_{ext}}{W(0)} + b_J +\alpha c_J \right)
\end{equation}
while the terminal velocity of the domain wall is
\begin{equation}
{\it v}_{T} \equiv v( \infty ) = - \frac{\gamma H_{ext}}{W(\infty)
\alpha}-\frac{c_{J}}{\alpha}.
\end{equation}
where $W(\infty)$ is the terminal wall width that is slightly smaller than
the initial N\'{e}el wall width $W(0).$
Equations (14) and (15) reveal the different roles played by the adiabatic
($b_J$ term) and non-adiabatic ($c_J$ term) spin torques: the adiabatic torque
is most important at the initial motion of the wall while the non-adiabatic 
$c_J$ controls the terminal velocity of the domain wall. 
The adiabatic torque causes the domain wall distortion. The distorted 
domain wall is able to completely absorb the adiabatic spin angular 
momentum so that the net effect of the adiabatic
torque on the domain wall velocity becomes null, i.e., domain wall 
stops. In contrast, the non-adiabatic spin torque behaves as 
a non-uniform magnetic field $c_J \partial {\bf M}/\partial x$ that can
sustain a steady state wall motion. 
Although the magnitude of the non-adiabatic torque $c_J$ is about
two orders of magnitude smaller than adiabatic torque $b_J$, the terminal
velocity is inversely proportional to the damping parameter which makes the 
velocity comparable to $b_J$. 

Finally, we emphasize that the present study has resolved an
outstanding mystery between the recent experimental observation \cite{Shinjo} 
and the theoretical prediction based on the
adiabatic spin torque. It has been recognized that a critical current
density of the order of $10^9-10^{10} A/cm^2$ is required to move
a perfect domain wall \cite{Tatara,Zhang} if we only use the adiabatic 
spin torque $b_J$. Experimentally, a velocity about 3 m/s was observed 
in a NiFe nanowire when a current density $1.2\times 10^8 A/cm^2$ was applied.
This velocity had been assumed to relate with $b_J$ \cite{Shinjo} in spite
of the apparent {\em qualitative and quantitative disagreement} between 
theory and experiment. Here, we have pointed out that $b_J$ is simply an
initial velocity of the domain wall and the measured velocity was the 
terminal velocity. For the experimental current density of
$1.2\times 10^8 A/cm^2$, the adiabatic spin torque alone is unable
to sustain a constant velocity. By including a
small non-adiabatic torque $c_J$, we find the domain wall velocity is now
$c_J/\alpha$ in the absence of the magnetic field, see Eq.~(15). 
Although the numerical values of both the exchange constant $J_{ex}$ and 
the damping parameter $\alpha$ are not precisely known in ferromagnets,
we estimate that the wall velocity should be $6 \sim 60$ (m/s) for the above
current density if we use the parameters indicated before (taking $\alpha
=0.01 \sim 0.1$ for permalloy). While the experimental velocity is smaller
than our estimated value, it is reasonable that we do not include any
defects that may reduce the observed velocity significantly.

The research was supported by NSF grants ECS-0223568 and DMR-0314456.


\begin{thebibliography}{1}
\bibitem{footnote}
The Walker's trial function breaks-down at a very high current density
at least $10^9 A/cm^2$, see Ref.~[8] and [13]. Here we assume the current 
is smaller than $10^9 A/cm^2$.
\end{thebibliography}
\end{document}